\documentclass[
 aps,
 prb,
 amsmath,
 amssymb,
 superscriptaddress
 ]{revtex4-2}

\usepackage{graphicx}
\usepackage{dcolumn}
\usepackage{bm}
\usepackage{chemformula}
\usepackage{siunitx}
 
\usepackage{xr}

\makeatletter
\newcommand*{\addFileDependency}[1]{
  \typeout{(#1)}
  \@addtofilelist{#1}
  \IfFileExists{#1}{}{\typeout{No file #1.}}
}
\makeatother

\newcommand*{\myexternaldocument}[1]{
    \externaldocument{#1}
    \addFileDependency{#1.tex}
    \addFileDependency{#1.aux}
}

\myexternaldocument{supportinginfo}

\begin{document}

\title{A semimetallic square-octagon (fes) two-dimensional polymer with high mobility}

\author{Tsai-Jung Liu}
\altaffiliation{Equal contribution}
\affiliation{Faculty for Chemistry and Food Chemistry, TU Dresden, Bergstrasse 66c, 01069 Dresden, Germany}

\author{Maximilian A. Springer}
\altaffiliation{Equal contribution}
\affiliation{Faculty for Chemistry and Food Chemistry, TU Dresden, Bergstrasse 66c, 01069 Dresden, Germany}
\affiliation{Helmholtz-Zentrum Dresden-Rossendorf, Institute of Resource Ecology, Permoserstrasse 15, 04318 Leipzig, Germany}

\author{Niclas Heinsdorf}
\affiliation{Institut f\"ur Theoretische Physik, Goethe-Universit\"at Frankfurt, Max-von-Laue-Strasse 1, 60438 Frankfurt am Main, Germany}

\author{Agnieszka Kuc}
\affiliation{Helmholtz-Zentrum Dresden-Rossendorf, Institute of Resource Ecology, Permoserstrasse 15, 04318 Leipzig, Germany}

\author{Roser Valent\'{i}}
\affiliation{Institut f\"ur Theoretische Physik, Goethe-Universit\"at Frankfurt, Max-von-Laue-Strasse 1, 60438 Frankfurt am Main, Germany}

\author{Thomas Heine}
\email[]{Thomas.Heine@tu-dresden.de}
\affiliation{Faculty for Chemistry and Food Chemistry, TU Dresden, Bergstrasse 66c, 01069 Dresden, Germany}
\affiliation{Helmholtz-Zentrum Dresden-Rossendorf, Institute of Resource Ecology, Permoserstrasse 15, 04318 Leipzig, Germany}
\affiliation{Department of Chemistry, Yonsei University, Seodaemun-gu, Seoul 120-749, Republic of Korea}

\date{\today}

\begin{abstract}
\noindent 
The electronic properties of $\pi$-conjugated two-dimensional (2D) polymers near the Fermi level are determined by structural topology and chemical composition. Thus, tight-binding (TB) calculations of the corresponding fundamental network can be used to explore the parameter space to find configurations with intriguing properties before designing the the atomistic 2D polymer network. The vertex-transitive \textbf{fes} lattice, which is also called square-octagon lattice, is rich in interesting topological features including Dirac points and flat bands. Herein, we study its electronic and topological properties within the TB framework using representative parameters for chemical systems. Secondly, we demonstrate that the rational implementation of band structure features obtained from TB calculations into 2D polymers is feasible with a family of 2D polymers possessing \textbf{fes} structure. A one-to-one band structure correspondence between fundamental network and 2D polymers is found. Moreover, changing the relative length of linkers connecting the triangulene units in the 2D polymers reflect tuning of hopping parameters in the TB model. These perturbations allow to open sizeable local band gaps at various positions in the Brillouin zone. From analysis of Berry curvature flux, none of the polymers exhibits a large topologically non-trivial band gap. However, we find a particular configuration of semimetallic characteristics with separate electron and hole pockets, which possess very low effective masses both for electrons (as small as $m^*_\mathrm{e} = 0.05$) and holes (as small as $m^*_\mathrm{h} = 0.01$). 
\end{abstract}

\maketitle
\section{Introduction}
 Topological materials have been intensively investigated in physical sciences~\cite{10.1103/RevModPhys.82.3045,10.1103/RevModPhys.83.1057, 10.1103/RevModPhys.88.021004}, because they exhibit numerous intriguing phenomena, such as the Quantum Spin Hall effect~\cite{10.1103/PhysRevLett.96.106802, 10.1126/science.1148047}, the Quantum Anomalous Hall effect~\cite{10.1103/PhysRevLett.61.2015, 10.1146/annurev-conmatphys-031115-011417}, topological superconductivity~\cite{10.1088/1361-6633/aa6ac7, 10.1038/s41567-020-0925-6}, or Majorana fermion excitation~\cite{majorana, 10.1038/nphys1380}, to mention a few. Similar to the case of graphene or the Lieb lattice~\cite{10.1038/nphys4080, 10.1038/nphys4105, PhysRevLett.114.245503, PhysRevLett.114.245504, 10.1126/sciadv.1500854}, the \textbf{fes} lattice is an ideal playground for studying topological quantum effects. Using tight-binding (TB) models, spin-density waves, superconductivity, and non-trivial topology were reported~\cite{10.1103/PhysRevB.101.224514,10.1088/1361-648X/ab1026,10.1140/epjb/e2019-100488-5,10.1103/PhysRevB.82.085106,10.1038/srep06918,10.1088/0953-8984/25/30/305602,10.1016/j.physleta.2017.12.051,10.1103/PhysRevB.98.245116,10.1103/PhysRevB.99.184506}. Additionally, with two different TB hopping parameters along the square and octagon edges, topological and magnetic phase transitions can be found~\cite{10.1103/PhysRevB.82.085106, 10.1038/srep06918,10.1016/j.physleta.2017.12.051}.
  
 The hypothetical realization of the \textbf{fes} lattice as a carbon allotrope, T-graphene~\cite{10.1103/PhysRevB.102.174509,10.1063/1.4828861,10.1103/PhysRevB.103.195104}, has been discussed but remains elusive, as such structures involve highly strained four-membered rings, and are not stabilized by an aromatic $\pi$-electron system. Highly crystalline conjugated two-dimensional (2D) polymers have become experimentally feasible with advanced surface synthesis techniques~\cite{10.1038/s41563-020-0682-z,10.1039/C2CS35157A, 10.1126/science.aan0202, 10.1021/acs.chemrev.9b00550}. Therefore, 2D polymers are an alternative to implement the \textbf{fes} net~\cite{10.1039/C9CS00893D}. For 2D polymers, the electronic structures around the Fermi level are often coarse-grained by TB models representing the respective network topology. This strategy can also be turned around and TB can be used to explore the space of electronic properties that are possible for 2D polymers~\cite{10.1021/jacs.8b09900}. Compared to ab initio calculations for atomistic structures, simple TB models can be tuned much more easily. Herein, we first use the TB model to explore the properties of the \textbf{fes} net, and, subsequently, propose a hypothetical conjugated 2D polymer with matching electronic structure.

 Being a conjugated carbon system, the spin-orbit coupling (SOC) in a 2D polymer will be small and results in tiny SOC-induced band gaps, similar in magnitude to the $\approx 1\,\si{\micro\electronvolt}$ in graphene~\cite{10.1103/PhysRevB.75.041401}. These tiny band gaps are too small for utilizing their topological properties at accessible temperatures, so additional mechanisms to increase the size of band gap are needed. Here, we present a chemically viable way to open local band gaps in a 2D polymer with \textbf{fes} network. Using a TB approach, we show that a band gap can be opened by tuning the relative strength of the hopping parameters while maintaining the space group. For the investigated system, however, whenever a new band gap opens due to structural perturbations in the TB model, the resulting system becomes topologically trivial. For certain configurations, a semimetallic state with separate electron and hole pockets arises. In semimetals, valence and conduction bands slightly cross the Fermi level at different points of the Brillouin zone and are not connected to each other. This leads to separate electron and hole pockets. Hence, the material can act as conductor with a limited number of electron and hole charge carriers. Besides topological semimetals (e.g. Weyl semimetals)~\cite{10.1021/acs.chemmater.7b05133}, topologically trivial semimetals were found to exhibit extreme magnetoresistivity~\cite{10.1103/PhysRevB.103.115119}.

  Secondly, we generate the corresponding \textbf{fes} net as a conjugated 2D polymer. It is shown that by varying the number of linker units - as they control the hopping strength between the vertices within the TB model (effectively the distance) - retains the electronic structure of the fundamental net.
  
\section{Methods}
 Our TB model Hamiltonian

\begin{equation}
H=\sum_{i}{\varepsilon_{i}c^{\dagger}_{i}c_{i}}+\sum_{\left\langle{i,j}\right\rangle}{t_{1}c^{\dagger}_{i}c_{j}}+\sum_{\left\langle\langle{i,j}\right\rangle\rangle}{t_{2}c^{\dagger}_{i}c_{j}}+H_{\text{SOC}}
\label{eq:H0}
\end{equation}

\noindent considers 1\textsuperscript{st}- ($t_1$) and 2\textsuperscript{nd}-neighbor ($t_2$) electron hopping parameters with on-site energies $\varepsilon_{i}$. Hopping parameters are usually negative ($t_i<0$), resulting in a node-less orbital between centres and stabilizing bond formation in chemical systems (see section~\ref{sec:tb_evaluations} of the Supporting Information (SI)~\cite{SI} for details). Since in 2D polymers electronic bonding between vertices is facilitated through the conjugated $\pi$ bonds rather than through the distance in space (see Fig.~\ref{fig:SI-distances})~\cite{10.1039/C9CS00893D}, only interactions (hopping parameters) through edges are considered. In order to account for SOC, analogously to the Haldane model, time-reversal symmetry breaking complex hopping terms were added (Eq.~\ref{eq:HSOC}). This induces local effective magnetic fluxes piercing the octahedron plaquettes~\cite{10.1103/PhysRevLett.61.2015,10.1103/PhysRevB.82.085106,mertz2019statistical}.

\begin{equation}
H_{\text{SOC}}=\sum_{\left\langle\langle{i,j}\right\rangle\rangle}{\lambda{\rm e}^{i\upsilon_{ij}\phi}c^{\dagger}_{i}c_{j}},
\label{eq:HSOC}
\end{equation}

\noindent where the staggered-flux parameter was chosen to be $\phi = \frac{\pi}{2}$; $\lambda$ is the SOC constant, $\upsilon_{ij}$ is defined as $d_{1} \times d_{2}$, where $d_{1}$ and $d_{2}$ are two normalized vectors denoting the connection from vertex $i$ to $j$; $\upsilon_{ij}$ can only be $\pm$1 and indicates the orientation path of an electron traveling from vertex $i$ to a 2\textsuperscript{nd}-neighbor vertex $j$ through $i$'s 1\textsuperscript{st} neighbor.

 For atomistic 2D polymers, band structures were evaluated using density-functional theory (DFT), employing the generalized gradient approximation by Perdew, Burke and Ernzerhof (PBE)~\cite{10.1103/PhysRevB.44.7888}, a double-$\zeta$ basis set~\cite{10.1080/00268970701598063} with SOC. A symmetric reciprocal grid with five \textit{k} points~\cite{10.1103/PhysRevB.56.13556}, as implemented in the AMS-BAND software package~\cite{10.1002/jcc.1056,BAND2019}, was used. Furthermore, the number of charge carriers in each pocket and topological properties of the polymer were evaluated. Hence, a TB model with parameters obtained from fitting the four characteristic \textbf{fes}-like bands around the Fermi level to the DFT band structures was employed.

\section{Results}
\subsection{Investigation of the fundamental \textbf{fes} network}
\begin{figure*}[h!]
	\includegraphics[width=\textwidth]{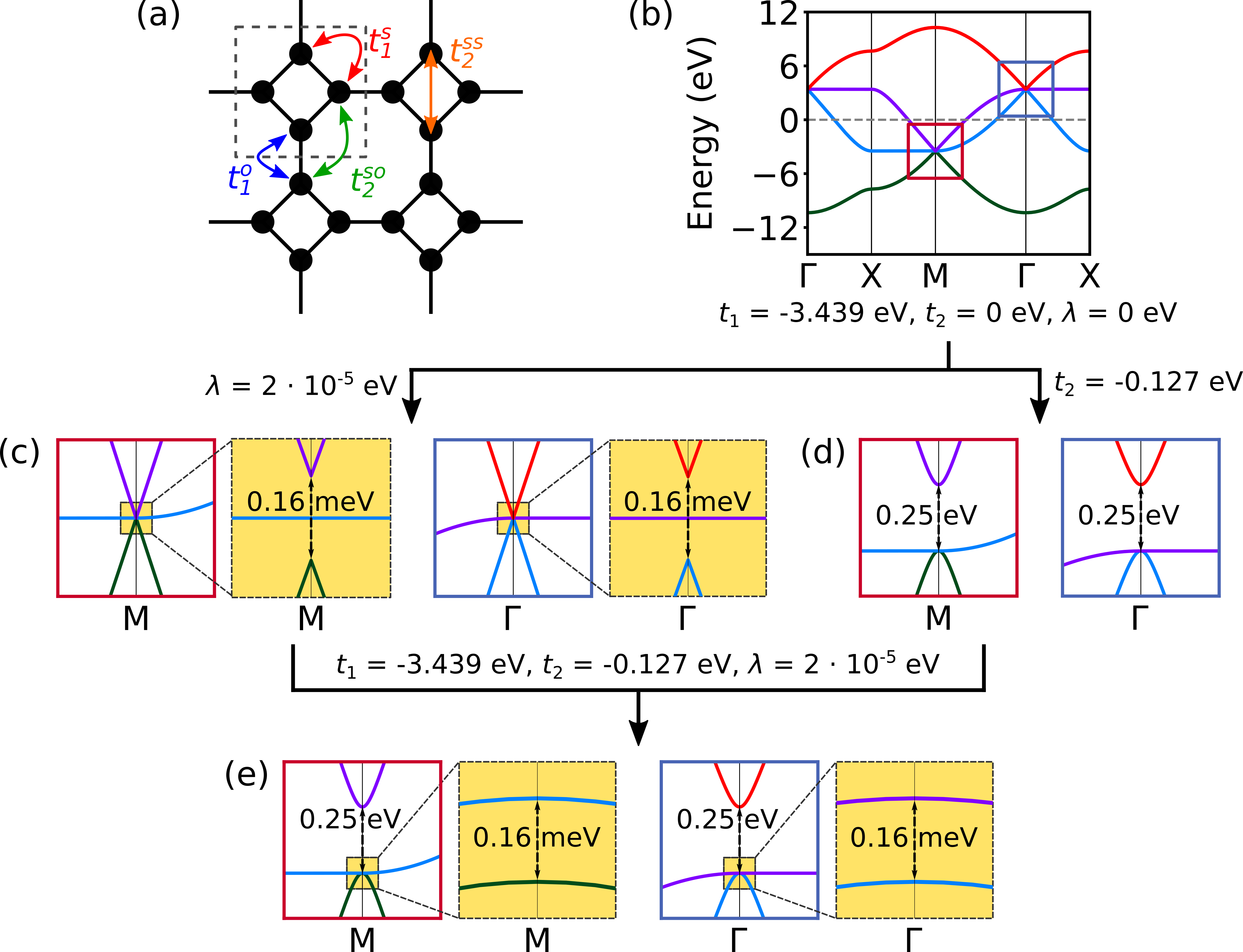}
	\caption{\label{fig:fes-neighbors} (a) The \textbf{fes} network with squares and octagons. Exemplary 1\textsuperscript{st}- and 2\textsuperscript{nd}-neighbor hoppings within the square, $t_1^s$ and $t_2^{ss}$, as well as between squares, $t_1^o$ and $t_2^{so}$, are shown. Arrows indicate which sites are involved in the interaction. (b) The corresponding \textbf{fes} band structure with only 1\textsuperscript{st}-neighbor hoppings ($t_1^o=t_1^s$). (c)-(e) Band structure details near the originally triply degenerate points at $\mathit{M}$ (red box) and $\mathit{\Gamma}$ (blue box) when other interactions are added to $t_1^o=t_1^s$, i.e., (c) 1\textsuperscript{st}-neighbor hoppings and SOC, (d) under consideration of 1\textsuperscript{st}- and 2\textsuperscript{nd}-neighbor hoppings, (e) 1\textsuperscript{st}- and 2\textsuperscript{nd}-neighbor hoppings, and SOC. Yellow boxes show small band gap openings due to SOC. Band structures are obtained from TB, for details on hopping parameters see SI section \ref{sec:tb_evaluations}~\cite{SI}.}
\end{figure*}

 For TB investigations of the fundamental networks, two symmetrically distinct types of first-neighbor hopping parameters can be used (see Fig.~\ref{fig:fes-neighbors}(a)): $t_1^s$ and $t_1^o$ for the 1\textsuperscript{st}-neighbor hopping within the square and between two squares, respectively. Analogously, the 2\textsuperscript{nd}-neighbor hopping parameters along two square edges ($t_2^{ss}$) and along one square and one octagon edge ($t_2^{so}$) are independent of each other. The band structure for the most commonly employed parameter choice~\cite{10.1103/PhysRevB.99.184506, 10.1103/PhysRevB.98.245116, 10.1103/PhysRevB.101.224514, 10.1088/1361-648X/ab1026, 10.1140/epjb/e2019-100488-5},  $t_1^o=t_1^s$, is shown in Fig.~\ref{fig:fes-neighbors}(b). The bands have a three-fold degeneracy at both the $\mathit{\Gamma}$ and $\mathit{M}$ points as there Dirac cones intersect a flat band (the corresponding Brillouin zone is shown in Fig.~\ref{fig:SI:bz})\cite{10.1103/PhysRevB.82.085106}.
 
\begin{figure*}[h!]
	\includegraphics[width=\textwidth]{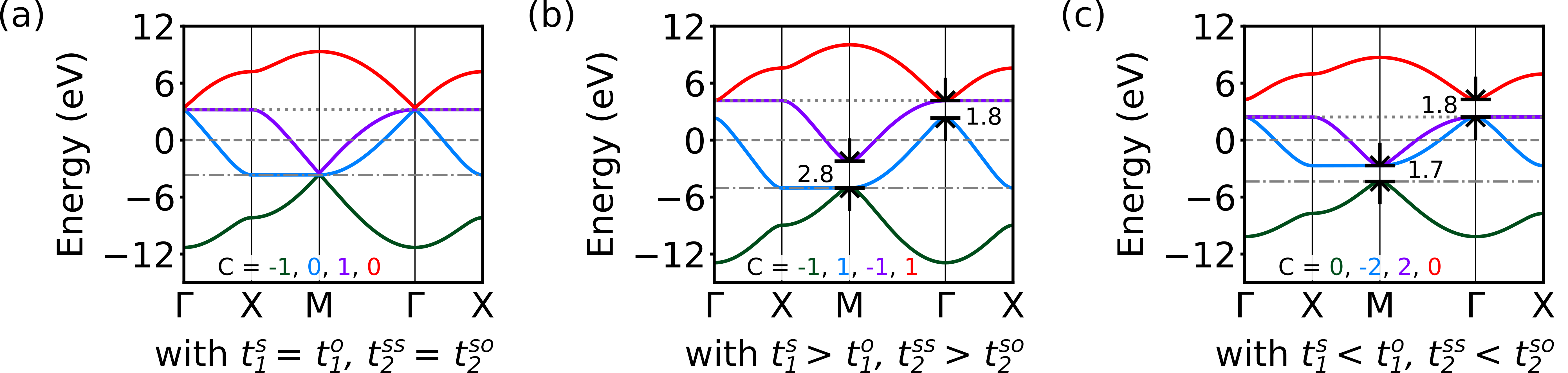} 
	\caption{\label{fig:fes-tune} The \textbf{fes} band structures and corresponding Chern numbers with different hopping parameters for intra-square and inter-square interactions. 
	The strength of SOC considered is $\lambda = 2 \cdot 10^{-5}~\si{\electronvolt}$: (a) $t_1^s = t_1^o = -3.438$~\si{\electronvolt},  $t_2^{ss} = t_2^{so} = -0.126$~\si{\electronvolt}, (b) $t_1^s = -3.438$~\si{\electronvolt}, $ t_1^o = -4.599$~\si{\electronvolt}, $t_2^{ss} = -0.126$~\si{\electronvolt}, $t_2^{so}= -0.185$~\si{\electronvolt}, and (c) $t_1^s = -3.438$~\si{\electronvolt}, $ t_1^o = -2.562$~\si{\electronvolt},  $t_2^{ss} = -0.126$~\si{\electronvolt}, $t_2^{so} = -0.086$~\si{\electronvolt}. Dash-dotted, dashed, and dotted lines indicate the Fermi level for one-, two-, and three-electron fillings, respectively.}
\end{figure*}

 Two effects that can lead to a formation of band gaps are considered: SOC and 2\textsuperscript{nd}-neighbor interactions. In Fig.~\ref{fig:fes-neighbors}(c), the effect of SOC is shown: a local band gap (in the order of $10^{-5}\,\si{\electronvolt}$) opens the formerly triply degenerate points at $\Gamma$ and $\mathit{M}$ points. Furthermore, there is the possibility to obtain a topologically non-trivial phase. Nevertheless, the opened gaps are too small for applications without further modification. Alternatively, sizable local band gaps of $0.25\,\si{\electronvolt}$ can be opened with 2\textsuperscript{nd}-neigbor interactions, as shown in Fig.~\ref{fig:fes-neighbors}(d). However, these gaps result in topologically trivial electronic structures. In Fig.~\ref{fig:fes-neighbors}(e), both effects are included. All degeneracies in the band structure are lifted. There are small topological band gap openings due to SOC ($E_g = 0.16~\si{meV}$) and larger trivial gaps due to 2\textsuperscript{nd}-neighbor hopping ($E_g = 0.25~\si{\electronvolt}$) at both $\mathit{\Gamma}$ and $\mathit{M}$ (cf.~Fig.~\ref{fig:fes-tune}(a)).

 In addition to 2\textsuperscript{nd}-neighbor hopping and SOC, the band degeneracy can be broken by structural perturbation. In organic polymer networks, this may be a promising and feasible strategy to complement the tiny SOC-induced band gaps. A similar idea has been successfully applied in carbon nanoribbons, where quantum confinement due to the finite ribbon width opens the band gap. This was first predicted theoretically~\cite{PhysRevLett.97.216803} and later confirmed experimentally~\cite{10.1038/s41586-018-0376-8, 10.1038/s41586-018-0375-9}.
 
 Apart from the common parameter choice $t_1^o=t_1^s$, hopping parameter configurations for unequal edges are possible. Fig.~\ref{fig:fes-tune} shows the \textbf{fes} band structures when considering both 1\textsuperscript{st}- and 2\textsuperscript{nd}-neighbor hopping parameters, as well as SOC. In addition, the Fermi level for different electron fillings is shown. Considering the original parameter choice $t_1^s = t_1^o$ (Fig.~\ref{fig:fes-tune}(a)), the system is a narrow-gap topological insulator (TI) for $\nicefrac{1}{4}$ filling (dash-dotted line), since the sum of Chern numbers of occupied bands is non-zero. However, for $\nicefrac{1}{2}$ filling (dashed line), it is a metallic system, whereas it is a trivial system for $\nicefrac{3}{4}$ filling (dotted line), since the sum of Chern numbers for occupied bands is zero. As shown in Fig.~\ref{fig:fes-tune}(b), hopping strengths between squares larger than within squares ($t_1^s > t_1^o$) open local band gaps between the 2\textsuperscript{nd} and 3\textsuperscript{rd} lowest band (blue and purple) at $\mathit{\Gamma}$ and $\mathit{M}$ points, resulting in a semimetallic system. 2\textsuperscript{nd} and 3\textsuperscript{rd} lowest band (blue and purple) cannot be fully separated by any filling. For the half-filled system (dashed line), the perturbed \textbf{fes} is a trivial semimetal with separate electron and hole pockets (around $\mathit{M}$- and $\mathit{\Gamma}$-points, respectively). In this hopping configuration, $\nicefrac{1}{4}$ (dash-dotted line) and $\nicefrac{3}{4}$ fillings posses nonzero Chern number. However, the SOC-induced gaps that are accessed with $\nicefrac{1}{4}$ and $\nicefrac{3}{4}$ fillings are very small (0.16~\si{meV} at both $\mathit{\Gamma}$ and $\mathit{M}$). In the opposite hopping configuration, considering smaller hopping strength along the octagon edges ($t_1^o > t_1^s$ in Fig.~\ref{fig:fes-tune}(c)), the degeneracy between highest and second highest bands (red and purple) at $\mathit{\Gamma}$-point and between lowest and second lowest bands (green and blue) at $\mathit{M}$-point are broken and large gaps open. However, the sum of Chern numbers of occupied bands are always zero when the Fermi level is placed at those gaps, and the two bands with nonzero Chern number again cannot be separated (blue and purple bands). Thus, their topological properties can not be exploited. While Chern numbers change with stronger SOC values (see Fig.~\ref{fig:SI:huge-soc} in SI), the necessarily large SOC cannot be found in metal-free, carbon-based systems.

\subsection{Implementation as 2D polymer}
 In the TB picture, the position of band gaps was tuned by alteration of relative strength of 1\textsuperscript{st}- and 2\textsuperscript{nd}-neighbor hopping parameters ($t_1^s $ vs.~$t_1^o$ and, hence, $t_2^{ss}$~vs.~$t_2^{so}$). This corresponds to different distances of vertices along the two distinct edge types in the 2D polymer. 2D polymers consist of connectors with three and linkers with two connection points to other building blocks. In order to implement the \textbf{fes} network as atomistic structure, one connector with three connection points to linkers (corresponds to vertices in Fig.~\ref{fig:fes-neighbors}a) and two types of linkers, a linear linker along the octagon edges and a bent one along the square edges, are necessary. By changing the chain length of one linker type and leaving the other unchanged, the relative interaction strength can be controlled. However, this analogy between TB and atomistic structure only works if the building blocks of the 2D polymer are fully $\pi$-conjugated.

\begin{figure}[t!]
\includegraphics[width=\textwidth]{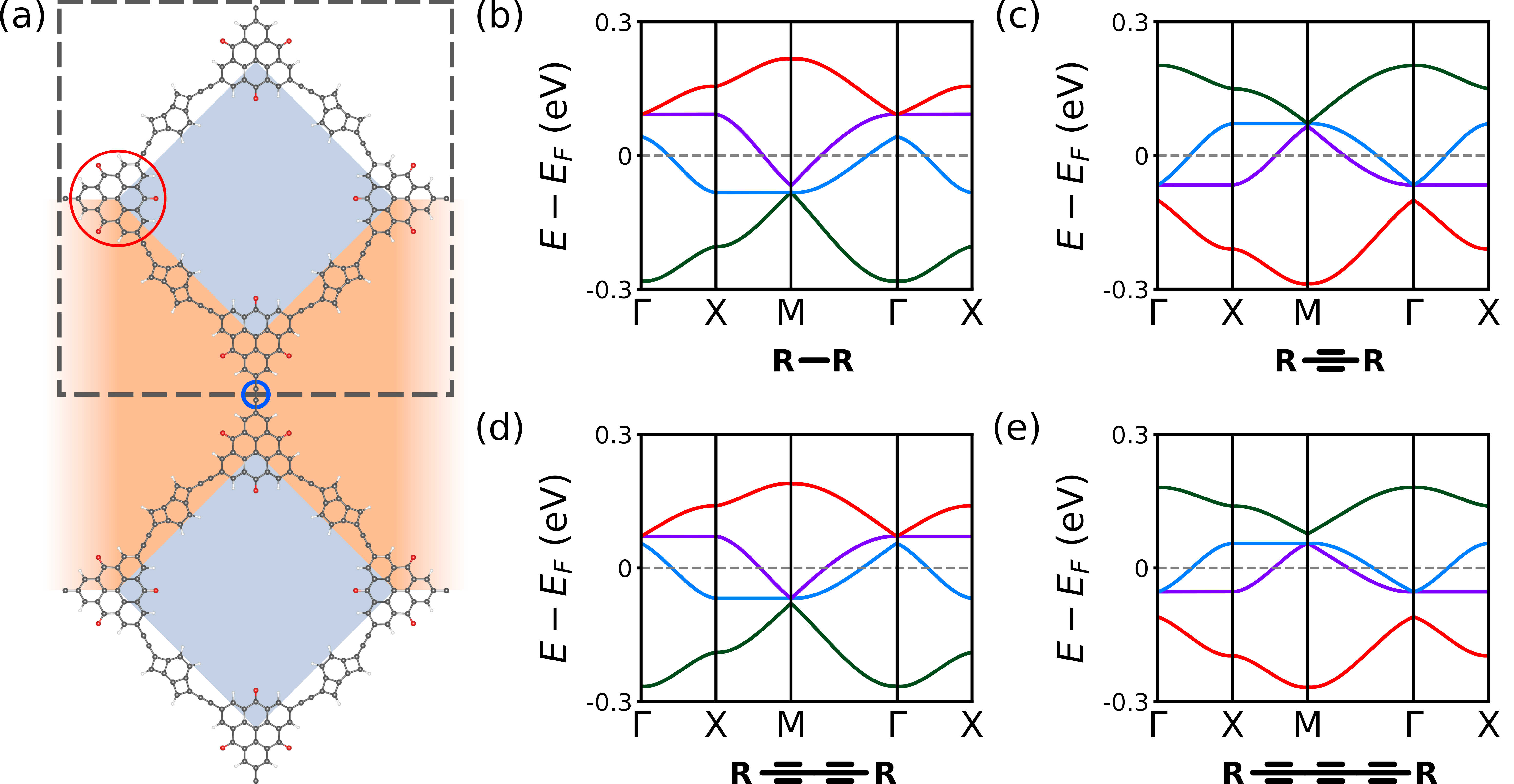}
\caption{\label{fig:atomistic-bands}(a) Atomistic structure of a hypothetical 2D polymer with the underlying \textbf{fes} net topology. Light blue square and orange areas indicate the position of the square and octagon of the \textbf{fes} net, respectively. Triangulene connectors (red circle) are put on vertex positions of the fundamental \textbf{fes} network, while two types of linkers are put on the square and octagon edges. The relative strength of linkers (effectively corresponding to relative strength of hopping parameters in the TB picture) can be controlled by leaving one linker unchanged and changing the length or the other linker. Here, the number of \ch{C2} units in the linear linker (blue circle) is changed to control its length. Hydrogen, carbon, and oxygen atoms are shown in white, grey, and red, respectively. (b-e) Band structures of the hypothetical 2D polymers with different number of \ch{C2} units as linkers connecting the squares from first-principles. The black dashed line indicates the Fermi level. For (b), the electronic effective mass is $m^*_\mathrm{e} = 0.01$ (blue band, $\mathit{\Gamma}$-point) and for holes is $m^*_\mathrm{h} = 0.05$ (purple band, $\mathit{M}$-point).}
\end{figure}

 Fig.~\ref{fig:atomistic-bands} shows one of the proposed 2D polymers (Fig.~\ref{fig:atomistic-bands}(a)) along with the corresponding band structures, obtained for different linker lengths between the square unit (Fig.~\ref{fig:atomistic-bands}(b-e)). Triangulene units (red circle in Fig.~\ref{fig:atomistic-bands}(a)), which were also used in the 2D polymer P\textsuperscript{2}TANG~\cite{10.1021/jacs.8b09900}, provide a singly occupied $\pi$ orbital. This resembles the situation in the TB calculations, where each site is treated as an in-plane centrosymmetric orbital (like the $p_z$-orbitals in graphene). Therefore, four triangulene units per cell represent the sites in the \textbf{fes} net. In order to form the characteristic squares of the \textbf{fes} net, triangulene units are connected with bent linkers (light blue squares in Fig.~\ref{fig:atomistic-bands}(a)). This feature is the same for all proposed 2D polymers. Thus, the distance of scattering centers within the square (coinciding with $t_1^s$ in the TB model) remains the same. The squares are either linked directly with each other using a \ch{C-C} bond between triangulene units or with \ch{C2} units (blue circle, corresponding to $t_1^o$ hopping according to Fig.~\ref{fig:fes-neighbors}(a)).

 Fig.~\ref{fig:atomistic-bands}(b-e) shows band structures of hypothetical 2D polymers with different number of \ch{C2} units used as linear linkers between the triangulene building blocks. In Fig.~\ref{fig:atomistic-bands}(b), the band structure of the polymer with a direct linkage between triangulene units (no extra linear linker connecting the squares) is shown. It coincides well with Fig.~\ref{fig:fes-tune}(b), where the interaction between squares is stronger than within squares ($t_1^s > t_1^o$ and $t_2^{ss} > t_2^{so}$). At the $\mathit{\Gamma}$ point, the third and fourth lowest bands (purple and red) are degenerate, whereas at the $\mathit{M}$ point, the lowest and second lowest bands (green and blue) are degenerate. The relative interaction strength is also reflected in the spatial distance between the scattering centers in the triangulene units, which is $21.3~\si{\AA}$ within the square and $10.3~\si{\AA}$ between the squares. Band structures for 2D polymers with one or two \ch{C2} units (Fig.~\ref{fig:atomistic-bands}(c,d)) are more similar to those shown in Fig.~\ref{fig:fes-tune}(a), where $t_1^{s} = t_1^{o}$ and $t_2^{ss} = t_2^{so}$. However, as all band structures for 2D polymers with an odd number of \ch{C2} units in the linear linker, the band structure in Fig.~\ref{fig:atomistic-bands}(c) seems to be flipped along the energy axis compared to the TB band structure of fundamental \textbf{fes}. The addition of $\pi$ orbitals leads to a change in the nodal structure of molecular orbitals for systems with an odd number of \ch{C2} units. It can be understood like a periodic 1D-chain of orbitals, where the addition of one orbital inverts the sign of the amplitude of the wave function. Hence, there is an overall sign change in the energy spectrum. The structure with three \ch{C2} units as linear linkers (Fig.~\ref{fig:atomistic-bands}(e)) is similar to Fig.~\ref{fig:fes-tune}(c), where $t_1^{s} < t_1^{o}$ and $t_2^{ss} < t_2^{so}$ due to the longer distance between the squares.
 
\begin{figure*}[t!]
    \centering
    \includegraphics[width=\textwidth]{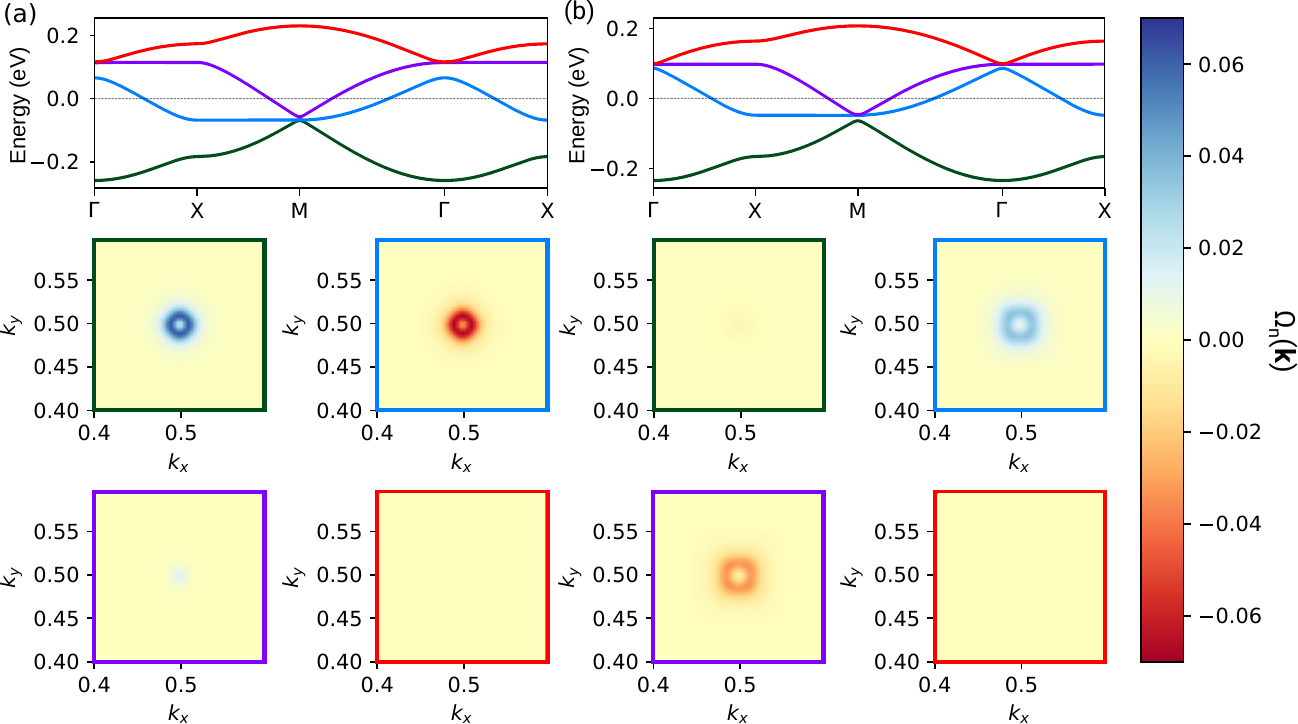}
    \caption{Extracted TB band structure and Berry flux $\Omega_n(\mathbf{k})$ around $\mathit{M}$ of the hypothetical 2D polymer with carbonyl bridge group. (a) without a linear linker between triangulene units (cf.~Fig.~\ref{fig:atomistic-bands}(b)) and (b) with two \ch{C2} units as linear linker (cf.~Fig.~\ref{fig:atomistic-bands}(d)). The colored frames of the Berry curvature plots indicate the corresponding band. In the structure without linear linker, the lower two bands (green and blue) form a Dirac cone at $\mathit{M}$, and consequently have an enhanced Berry curvature flux in the vicinity of that point. Albeit very close in energy, the third band (purple) exhibits almost no Berry curvature flux. By inserting the linker (b), Berry curvature flux is transported to higher lying bands in the spectrum, bringing it closer to the Fermi level. See Tab.~\ref{tab:SI-fit-C} in the SI~\cite{SI} for TB parameters.}
    \label{fig:berryFlux_transport}
\end{figure*}

 In order to investigate the topological properties of the proposed 2D polymers, a TB model was extracted from each ab initio band structure (see SI section~\ref{sec:tb-fitting} for details). Exemplary for the other structures, the 2D polymers with carbonyl bridge group without linear linker (cf.~Fig~\ref{fig:atomistic-bands}b) and with two \ch{C2} units in the linear linker (cf.~Fig~\ref{fig:atomistic-bands}d) are assessed. In Fig.~\ref{fig:berryFlux_transport}, the Berry curvature flux $\Omega_n(\mathbf{k})$ for each band is shown (see SI section~\ref{sec:berry} for details). For the polymer without linear linker, there is a strong enhancement of $\Omega_n(\mathbf{k})$ on the two lowest bands (green and blue) close to the $M$ point. These fluxes have opposite signs and behave very similar to the case of the Haldane model on the \textbf{hcb} lattice at $K$ and $K^{'}$. The third band (purple), which is very close in energy, carries almost no $\Omega_n(\mathbf{k})$. As shown in Fig.~\ref{fig:berryFlux_transport}(b), inserting two \ch{C2} units as a linker pushes the two middle bands (blue and purple) slightly upwards, creating a band gap to the lowest bands (green). Additionally, and more importantly, there is a strong redistribution of $\Omega_n(\mathbf{k})$ to the second highest band (purple), which previously carried almost no contribution at all. As a consequence, the Dirac cone sits higher in energy, not because of an overall shift in energy, but because of a transfer of $\Omega_n(\mathbf{k})$ to bands that naturally occupy states closer to the Fermi level. Topological features in the bulk only enforce topologically protected edge states if they sit directly at or very close to the Fermi level. Thus, this feature is more accessible without any external field or charge carrier doping. However, if chemical alteration is too drastic, the Berry curvature flux transfer can become obstructed (cf. SI section \ref{sec:obstructberry}~\cite{SI}).
 
 Since there is a one-to-one correspondence of TB model and 2D polymer, the system with stronger interaction between squares than within squares (Fig.~\ref{fig:fes-tune}(b) and Fig.~\ref{fig:atomistic-bands}(a)) is semimetallic. Since both the electron and the hole pocket are remains of the Dirac points in the unperturbed system, the effective masses are remarkably small. For the system shown in Fig.~\ref{fig:fes-tune}(b), very small effective masses ($m^*_\mathrm{e} = 0.05$ for the blue band at $\mathit{\Gamma}$-point and $m^*_\mathrm{h} = 0.01$ for purple band at $M$-point) were found. For the 2D polymers with other bridge groups, small effective masses were calculated, as well (see Tab.~\ref{tab:SI_effective_masses_CH2}-\ref{tab:SI_effective_masses_O}). The largest effective mass was found for the polymer with methylene bridge group and no extra linear linker, which were $m^*_{\mathrm{e}} = 0.06$ and $m^*_{\mathrm{h}} = 0.05$. These are remarkably low effective masses for 2D polymers. This kind of low effective mass can only be found for transport between layers~\cite{10.1021/acs.jpcc.5b11928}. However, it is limited to interlayer transport. Within single layers, the effective masses are normally much larger in the $10^{-1}$ order of magnitude~\cite{10.1021/acs.chemmater.8b04986,10.1039/C6CP06101J,10.1021/acs.jpcc.5b11928}, and can easily go up to the order of $m^* = 1$ or even larger~\cite{10.1039/C6CP06101J,10.1021/acs.jpcc.5b11928,10.1021/acsami.8b14446,10.1021/jacs.9b07644}. The suggested 2D polymers possess outstanding low effective masses, pointing towards very good electron transport properties.
 
 Using the fitted TB model, the number of states in electron- and hole-like pockets were calculated for the semimetalic 2D polymers. Since the $\mathit{\Gamma}$- and $M$-points are symmetry connected in the fundamental net, the charge carriers compensate well (cf.~Fig.~\ref{tab:SI_charge_carriers}). E.g., for the polymer shown in Fig.~\ref{fig:atomistic-bands}(a), the number of states in the electron-like pocket at the $\mathit{M}$-point and the hole-like pocket at the $\mathit{\Gamma}$-point is in each case 0.25 states per unit cell. This points towards a potential use in magnetoresistive materials, with the extreme magnetoresistive effect being associated with electron-hole compensation~\cite{10.1103/PhysRevB.103.115119}. For longer linear linkers, the number of states in the pockets increases compared to the system without linear linkers with the same bridge group. The largest difference of states in electron- and hole-like pockets is found for the methylene bridge group and four \ch{C2} units as linear linkers with 0.25 and 0.20 states per unit cell, respectively.
 
\section{Conclusions}
 We explored the electronic properties of the fundamental \textbf{fes} net (square-octagon net) using a TB approach. It was shown that the band structure and topological properties can be tuned by changing the relative strength of hopping parameters of two distinct edges. Furthermore, different topological states of the system (topological insulator, topological metal) can be accessed by changing the electron filling. However, the SOC-induced band gaps are too small for practical use. Based on the TB investigations, a hypothetical atomistic 2D polymer with \textbf{fes} net is proposed. By varying the length of linkers (representing edges in the material implementation), the band structure can be tuned in a one-to-one correspondence to the TB results. We showed how the Berry curvature is reconstructed whenever a sizeable band gap is opened due to structural perturbation with a TB model extracted from ab initio band structures. For the configuration with stronger interaction between squares than within the squares, a semimetallic state with remarkably small effective masses can be found. The proposed 2D polymers serve only as representative example for implementation, as many other structures may be possible to realize. They give the possibility to explore properties of the \textbf{fes} net via structural modifications. Such a proposal is, thus, much closer to the experimental realization than the models published so far.

\begin{acknowledgments}
 M.S.~acknowledges Dr.~Thomas Brumme, Dr.~Patrick M\'{e}lix and Dr.~Miroslav Polo\v{z}ij for fruitful discussions. Financial support by the International Max Planck Research school and by Deutsche Forschungsgemeinschaft (CRC 1415 and PP 2244) is acknowledged. The authors thank ZIH Dresden for the use of computational resources. N.H.~and R.V.~were financially supported by the Stiftung Polytechnische Gesellschaft Frankfurt and by the Deutsche Forschungsgemeinschaft (DFG, German Research Foundation) through TRR 288 - 422213477 (project A05, B05).
\end{acknowledgments}
\end{document}